# Compton shift and de Broglie frequency

Raji Heyrovská

J. Heyrovský Institute of Physical Chemistry, Academy of Sciences of the
Czech Republic, Dolejskova 3, 182 23 Prague 8, Czech Republic.
Raji.Heyrovska@jh-inst.cas.cz

**Abstract.** Compton scattering is usually explained in terms of the
relativistic mass and momentum. Here, a mathematically equivalent and
simple non-relativistic interpretation shows that the Compton frequency
shift is equal to the de Broglie frequency associated with the moving
charged particle (e.g., an electron). In this work, the moving electron is
considered as a particle and the electromagnetic energy associated with it is
shown to be proportional to the de Broglie frequency. This energy is
released when its motion is arrested, as for example on a diffraction screen,
where it causes the observed interfernce patterns. Thus, electrons transport
electromagnetic energy from a source to a sink.

An introduction to Compton effect can be found in many books, e.g., see
[1]. A bibliography on the various interpretations using relativistic and
quantum theories can be found in [2]. Here, a simple non-relativistic view of
the phenomenon is presented.

Essentially, in Compton scattering, an incident X-ray of wavelength

$\lambda_{in}$ interacts with matter and the scattered ray has a longer wave length $\lambda_\theta$, the value of which depends on the scattering angle $\theta$ from the incident ray. In the case of interaction with electrons, the Compton wavelength shift $\Delta\lambda_{e,\theta}$ is given by [1],

$$\Delta\lambda_{e,\theta} = \lambda_\theta - \lambda_{in} = \lambda_{C,e}(1- \cos \theta) \qquad\qquad (1)$$

where $\lambda_{C,e}$, the Compton wavelength (obtained as the Compton shift for $\theta = \pi/2$) is found to be,

$$\lambda_{C,e} = (h/m_e c) \ [= (I/m_e) = 2\pi(e^2/\kappa v_\omega)(1/m_e c)] \qquad\qquad (2)$$

h is the Planck constsnt and $m_e$ is the rest mass of the electron. The equivalent terms introduced here in the square brackets consist of I = (h/c) (= $\lambda_{C,e} m_e$), the moment of inertia of the electron with respect to a point at a distance $d_e = \lambda_{C,e}$, $\kappa = 4\pi\varepsilon_o$ (where $\varepsilon_o = 1/\mu_o c^2 = 1/Zc$ is the electric permittivty, $\mu_o$ is the magnetic permeability and Z is the impedance of vacuum), $(e^2/\kappa v_\omega) = \hbar = h/2\pi$, the angular momentum of spin is the product of charge (e) and magnetic pole strength $(e/\kappa v_\omega)$ of the spinning unit charge, $v_\omega = \alpha c$, is the velocity of spin (due to the electromagnetic torque arising from the mutually perpendicular electric and magnetic fields) and $\alpha$ is the fine structure constant.

The energy lost by the incident X-ray is, as per relativity theory [1], due to a gain in the mass $(\delta m)_\theta$ of the (recoil) electron, whereas here it is explained in (mathematically equivalent) terms of the gain in the linear velocity, $v_{e,\theta}$ of the electron.

The equations for the conservation of energy and momentum are given respectively by,

$$h(\nu_{in} - \nu_\theta) = h\Delta\nu_{e,\theta} = (\delta m)_\theta c^2 = m_e(\gamma_\theta - 1)c^2 = m_e c v_{e,\theta} \qquad (3)$$

$$(h/c)\Delta\nu_{e,\theta} = h[(1/\lambda_i) - (1/\lambda_\theta)] = (\delta m)_\theta c = m_e(\gamma_\theta - 1)c = m_e v_{e,\theta} \qquad (4)$$

where $\nu_{in}$ and $\nu_\theta$ are the frequencies and $\gamma_\theta$ is the relativity factor. The last terms in equations (3) and (4) imply that an electron initially at rest with velocity $v_o = 0$ (or having a constant initial velocity, $v_o$) gains the velocity $v_{e,\theta} = c(\gamma_\theta - 1)$ at the expense of the energy $h\Delta\nu_{e,\theta}$ lost by the incident X-ray.

The wavelength $\lambda_{dB,e,\theta}$, frequency $\nu_{dB,e,\theta}$ and period $\tau_{dB,e,\theta}$ of the de Broglie wave of the electron in motion are given by,

$$\lambda_{dB,e,\theta} = (h/m_e v_{e,\theta}) = c/(\Delta\nu_{e,\theta}) = c\lambda_{C,e}/v_{e,\theta} \qquad (5)$$

$$\lambda_{dB,e,\theta}/c = 1/\nu_{dB,e,\theta} = 1/\Delta\nu_{e,\theta} = \lambda_{C,e}/v_{e,\theta} = \tau_{dB,e,\theta} \qquad (6)$$

$$\nu_{in} = \nu_\theta + \nu_{dB,e,\theta} \qquad (7)$$

These equations show that the de Broglie wave is an electromagnetic wave of velocity c and that the Compton frequency shift ($\Delta\nu_{e,\theta}$) is equal to the frequency $\nu_{dB,e,\theta}$ of the de Broglie wave. The energy $h\nu_{dB,e,\theta}$ of the de Broglie wave is the electromagnetic energy associated with the moving electron, which it received from the incident electromagnetic wave.

Note that when the moving electron is brought to rest (or to its original velocity $v_o$), the electromagnetic energy will be released by the electron. Thus, a moving electron transports electromagnetic energy from a source to a sink. In diffraction experiments such as those by Davisson and Germer mentioned in [1] as a demonstration of the wave nature of electrons, actually electrons emitted from the source lose their momentum on coming to a rest on the screen and the electromagnetic energy released appears in the form of waves which cause the interference pattern. Therefore, note that the present work shows that electrons can be treated as particles and not as waves.

It also follows from equations (2) and (6) that,

$$\lambda_{C,e} = d_e = v_{e,\theta}\tau_{dB,e,\theta} = h/m_e c = I/m_e \qquad (8)$$

where $d_e$ is a fixed distance and is the product of the velocity $v_{e,\theta}$ and the period $\tau_{dB,e,\theta}$ of one de Broglie wave. If $v_\omega$ is considered as the spinning velocity acting on the distance $d_e$, the angular velocity of spin $\omega_e$ and the corresponding energy are given respectively by,

$$\omega_e = v_\omega/d_e = \alpha c/\lambda_{C,e} = \alpha m_e c^2/h \qquad (9)$$

$$\hbar\omega_e = \hbar v_\omega/d_e = \hbar\alpha c/\lambda_{C,e} = \alpha m_e c^2/2\pi \qquad (10)$$

**Acknowledgement:** This work was financed by Grant 101/02/U111/CZ